\documentclass[
,twocolumn
]{article}
\usepackage{hyperref,graphicx,url}

\begin{document}


\title{Design and Test of an adaptive augmented reality interface to manage systems to assist critical missions.}

\author{Dany Naser Addin and Benoît Ozell\\
 \small \texttt{\{dany.naser-addin,benoit.ozell\}@polymtl.ca}\\
 \small Department of Computer and Software Engineering, \\
 \small Polytechnique Montréal, Québec, Canada}

\maketitle



\begin{abstract}
We present a user interface (UI) based on augmented reality (AR) with head-mounted display (HMD) for improving situational awareness during critical operation and improve human efficiency on operations. The UI displays contextual information as well as accepts orders given from the headset to control unmanned aerial vehicles (UAVs) for assisting the rescue team. We established experiments where people had been put in a stressful situation and are asked to resolve a complex mission using a headset and a computer. Comparing both technologies, our results show that augmented reality has the potential to be an important tool to help those involved in the emergency situation.
\end{abstract}

Keywords:
\texttt{augmented reality}, \texttt{interactions}, \texttt{autonomous system}, \texttt{drone swarm}, \texttt{situation awareness}, \texttt{critical situations}, \texttt{firefighters}



\section{Introduction}

During critical situations, many humans risk their lives to save others. Relying more on technology can alleviate such risk and enhance the emergency response success rate. Previous works implemented unmanned aerial vehicles (UAVs) based systems to prevent human from accessing dangerous areas themselves.

Information and communication technology recent developments have made data more accessible than ever, challenging human capacity to quickly process massive loads of information. In case of emergency response for instance, being able to process information in time leads to risk reduction and success rate increase. We propose to use Augmented Reality (AR) to improve such system and provide better environment visualization and assets control.

AR is a technology that superimposes virtual information on the environment and is becoming more and more common in our daily lives. The advanced development of mobile phones available to the public makes it possible to provide numerous applications in AR or virtual reality (VR). Today, augmented and virtual reality has become accessible to everyone through the use of HMD, which allows the use of AR from a different approach. These visual representations are displayed on top of the real environment to provide additional information with the aim to assist people in complex tasks or for simple entertainment purposes.

This article is divided into three sections. First, we present emergency situations and the concept of situational awareness. We also review teleoperation with mixed reality and UAV-based rescue system. Second, we explain details on our AR-based UI design and present our implementation. Third, we present the tests conducted and discuss the results regarding the impact on humans' stress in the emergency response.

This work was carried out during the covid pandemic. Some adjustments in the experimentation have been set up . The experiment was to initially involve firefighters in order to use their knowledge of the emergency environment to evaluate the augmented reality technology in emergency situations. As a substitute, students from Polytechnique Montréal made the experiment without having any emergency's knowledge. The experimentation has been revised in order to evaluate the effectiveness of AR in emergency situations with these people.

\section{Related work}\label{related_work}

\subsection{Emergency Response’s Strategies}

The arrival of a natural disaster or an attack alert can represent different cases of an emergency operation. Many actions are carried out under considerable stress for rescue or military missions. It has been shown that 71\% of aviation accidents are due to human error~\cite{lee_effect_2014}. These errors are the result of high workloads and high levels of stress. Researchers report that reducing the operator’s workload and information to be processed has a positive impact on stress. On the contrary, if a person doesn’t have enough information about the situation, both mental and physical workloads become too great and this generates frustration which leads to errors. The term “Situational awareness” was introduced in 2003 by Endsley et al.~\cite{endsley_designing_2003} as the understanding of what is happening around us, the current situation and the one to come. The concept also regroups the filtering of important information to perform tasks at hand. Stress as well as a highly physical and mental workloads are factors that restrict situational awareness. This distracts the individual and prevents him or her from making the right decisions. In emergency situations, actors need to be careful about their situational awareness.

To measure the situational awareness of a person, Endsley et al. created the Situation Awareness Global Assessment Technique (SAGAT)~\cite{endsley_situation_1988} which measures the situational awareness of pilots in flight operations. Questions are asked to the pilot during the mission to measure the stress, they explained that humans tend to generalize if their feeling is asked at the end of the test which prevents a correct assessment. They use immersive technologies to design exercises for emergency responders and test usability with three immersive displays: a simple screen, a giant curved screen and a VR HMD. Each of these technologies has very different properties and each of them can be useful depending on the scenario. For training firefighters, Clifford et al. found out two important limitation factors: cost and environmental impact~\cite{clifford_development_2018}. Another method created was the NASA-TLX form~\cite{nasa_nasa-tlx_2019} consists of a set of questions that requires one to scale one’s effort from 0 to 100 in several types of mental and physical actions. Lee et al. set up a simulation of a helicopter attack to evaluate how the pilots feel~\cite{lee_effect_2014}. They obtained results using this NASA-TLX to evaluate the physical and mental loads and frustration related to factors to evaluate situational awareness.

An autonomous system (AS) is an Unmanned Aerial Vehicles (UAVs) which are robotic machines acting under human control. UAVs operate synchronously to complete tasks. Drones are from the UAV group and Yuan et al. listed numerous use cases~\cite{yuan_ultra-reliable_2018}. Research found various possibilities with UAV and several sensors, including cameras depending on the situation (video, surveillance, etc.)~\cite{noor_remote_2018}. An autonomous system poorly handled by human supervision can degrade the efficiency of using the machines~\cite{atyabi_current_2018}. Atyabi et al. present the necessity that a stand-alone system is not to become totally independent of any form of directive, but for that it must significantly reduce the mental and physical workloads of the user to improve situational awareness. In order to manage UAVs, people use a control station to optimally manage several air or ground vehicles as well called as a mission planner~\cite{brumitt_dynamic_1996}. It allows several objectives to be accomplished in dynamic environments. It is an essential element in the progress of an emergency operation. Reducing the human presence in dangerous areas impacts on the human risk, which can be done by using autonomous systems. Carrillo-Zapata et al. found that using a robotics swarm is really efficient in fire and rescue fields~\cite{carrillo-zapata_mutual_2020}. In a mutual work with firefighters, they found the swarm is improving performance if the data provided by the swarm is relevant and it is important to let the decision-making to humans.

\subsection{Mixed Reality Technologies}

\begin{figure}[ht]
	\centering
	\includegraphics[scale=0.36]{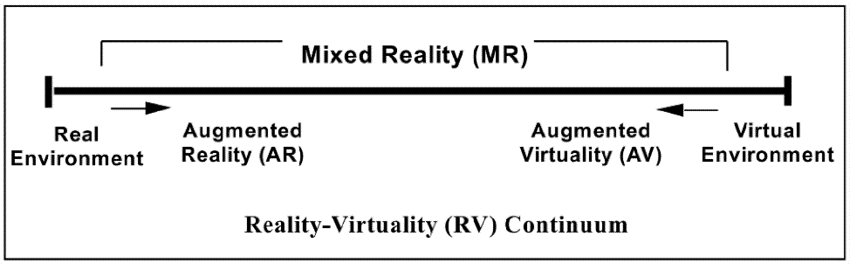}
	\caption{Virtual Reality and Augmented Continuum Milgram}
	\label{figure:continuum}
\end{figure}

Milgram defined the concept of AR as the overlaying of visual information into the real world~\cite{milgram_augmented_1995}. Display of a hologram in a real room corresponds to the left side of the scale (figure~\ref{figure:continuum}), considering it’s based mainly on a real environment. Conversely on the right part of the scale, VR cuts off all interaction from the real world to immerse users in a different and synthetic reality. The concept of “\textit{Mixed reality}” (MR) at the centre of the scale repose of using information from real environment (room configuration, use of real objects) to display digital information.

Navigation through AR implies the use of interfaces, defined by Hookway~\cite{hookway_interface_2014} as a bridge between human experience and allowing actions that it can’t undertake alone. Companies such as Microsoft~\cite{microsoft_conception_2019} or Apple~\cite{apple_design_2019} designed their own “user-friendly” AR interfaces for support like a computer or mobile device. Their solution is based on a 2D interface (e.g., touch screen) whereas AR using HMD includes the use of the third dimension. Argenta et al. designed interface to support tactical (military) missions and consider gestures, voice and touch with AR~\cite{argenta_graphical_2010}. Bertrand et al. reviewed this change of dimensions with common interface principles like e-mail applications~\cite{bertrand_augmented_2018}, they made adjustments to perform a pleasant and intuitive interface in 3D AR environment.

For interactions in AR, the first design of the augmented reality headset using the fingers of the hand as an interaction tool was made by Goos~\cite{goos_augmented_2000}. It was one of the first researchers to submit the prototype of an AR HMD. Their design consists of a user menu displayed on the hand, superimposing a button on each of the fingers. The user can then select by placing the index finger of the other hand on a button. Today, we can interact in AR using emerging technologies with headsets in different ways. By using the gaze to select elements~\cite{ajanki_augmented_2011}, recent AR headsets now support voice recognition and the tracking of user’s hands and eyes with Hololens 2~\cite{microsoft_hololens_2019} and Magic Leap 1~\cite{magic_leap_magic_2019}. However, the field of view of HMD remains limited today. People can now create holographic content on a computer and stream it directly to a HMD via network by Park et al.~\cite{park_volumetric_2018}. The latter proposed the idea as a solution to relocate computations and so enhance performance.

Das and his or her partners worked on a very complete system for improving the situational awareness of military agent to analyze environments using a swarm~\cite{das_rapid_2018}. Unfortunately, they just mentioned AR for simple visualization and didn’t use AR for the improvement of situational awareness. In another work, the same authors used a headset sensor to control robots~\cite{das_mixed_2017}. They detect the movements of hands to place passage points to guide robots. Their objective was to design a human-machine interface adapted in mixed reality to control a machine architecture. The interface wasn’t designed for emergency environment. Cousins et al. and Brizzi et al. interacted on AR for teleoperations with different arm robots~\cite{cousins_development_2017}~\cite{brizzi_effects_2018}. With the improvement of AR, Trzcielinski et al. found that humans learned faster to control the robotic arm thanks to augmented reality technology using a HMD~\cite{Trzcielinski_advances_2012}. The teleoperation was adapted here for learning and not for use in concrete situations. Patel et al. researched about controlling a swarm of small robots using AR with a tactile computer~\cite{patel_mixed-granularity_2019}~\cite{patel_improving_2020}. They made a comparison between robot-oriented and environment-oriented methods using AR with one or multiple participants at the same time.

Looking at the research on Mixed Reality technologies and emergency operation, we observed that some of AR applications were made and adapt for teleoperation, but not for users in emergency situations and not designed to reduce stress load. HMD nowadays can be efficient to improve humans situational awareness and our goal is to know if AR can help during emergency cases controlling a swarm of drones. Is AR using HMD capable of assisting humans during an emergency operation? Knowing that the field of emergency regroups various types of situations, we decided to focus on firefighters operation in fire in a High-Rise Building (HRB). Recently Sullivan et al. developed a first AR application using the functionalities from the headset Magic Leap 1 for managing a swarm~\cite{sullivan_augmented_2020}. They explored the methods of giving orders to the swarm and didn't focus of the data provided by the swarm.

\section{Methodology and Experimentation}
\subsection{Firefighters of Montréal (SIM)}\label{SIM}
Conan~\cite{conan_projet_2019} surveyed and analyzed the habits and procedures of professional firefighters of Montréal. In particular, three main problems have been identified in case of a high-rise building (HRB) intervention. For each problem, some important information is referenced in the table~\ref{table:question_firefighters_hrb}.

\begin{enumerate}
	\item \textbf{P1 - Location at all times:} Location is generally one of the critical and fundamental points of the emergency services, especially for firefighters where there is a lack of visibility between the coordinator and all the firefighters during a fire.

	\item \textbf{P2 - Evolution of the flames:} Obtaining the maximum amount of information about an emergency operation is important for the progress of the mission. This includes the visualization of the situation from various angles.

	\item \textbf{P3 - Optimization of understanding:} In emergency situations, operations generate information to process for humans. Therefore this implies a consequent information processing to facilitate the understanding of the situation in which firefighters find themselves.
\end{enumerate}

\begin{table}
\begin{center}
  \caption{Firefighters’ questions on fire emergency in HRB}
  
\begin{footnotesize}
\begin{tabular}{|c|p{0.7\linewidth}|}
  \hline
  \footnotesize\textbf{Field} & \footnotesize\textbf{Questions}\\
  \hline
  P1 & Where are the firefighters in the danger zone?\\
  \hline
  P2 & Where’s the seat of the fire?\\
  & How will the intensity of the fire progress in a short time? (louder/weaker)\\
  \hline
  P3 & How many human and animal casualties are there, and where are they?\\
  & What are the different needs of :
  \begin{enumerate}
	\item water
	\item man
	\item material
	\item aerie (air)
  \end{enumerate}\\
  & What is the situation in the risk zone?
  \begin{enumerate}
	\item Where are the load-bearing walls?
	\item What are the building materials used?
	\item What is the layout of the pipes?
	\item What is the plan of the area?
	\item Where are the emergency exits?
  \end{enumerate}\\
  \hline
\end{tabular}
\end{footnotesize}
\label{table:question_firefighters_hrb}
\end{center}
\end{table}

\subsection{Objectives}\label{Objectives}
Our goal is to evaluate the effectiveness of AR with the use of a HMD during a virtualized fire operation in a HRB. From our literature review in the previous section~\ref{related_work}, we studied these hypotheses:

\begin{enumerate}
\begin{footnotesize}
\item H\textsubscript{a1} - This interface in AR allows for a better understanding of information in the field of emergency response.
\item H\textsubscript{a2} - This interface in AR improves situational awareness.
\item H\textsubscript{a3} - This interface in AR makes decision-making easier.
\item H\textsubscript{a4} - This interface in AR makes it easier for mission objectives completion.
\end{footnotesize}
\end{enumerate}

First we understood the needs of Montréal's firefighters (SIM). Then we designed an AR application managing swarm drones for critical situations based on our research. To support our assumptions, we also designed a scenario for simulation with a fire in HRB. We confronted technology including an AR application for HMD with a simple computer software to manage a swarm in stressful situations. We tested our scenario case with people of different kinds of ages and experience to get their feedback about our work. This section contains the needs of firefighters in real case, the conception of our application, the designed simulation scenario and methods for evaluating our experience.

Regarding the questions mentioned in the section~\ref{SIM}, our technology uses an ad hoc network, it is possible to geolocate (without the presence of mobile networks or internet) all the electronic tools that can be either attach on robots or present on humans to obtain in real time the position of each. In order to understand the whole situation and to act quickly, UAVs composing the autonomous system are technologies with great ease of movement that allow to quickly photograph and map the area. Managing this data then displaying it on a 3D and 2D medium must increase knowledge of the issues in the field.

\subsection{Proposed Technology}\label{technology_solution}

We created an AR application where the user controlled an AS composed by drones. These drones have cameras for mapping. Each of these UAVs achieves a common objective, they evaluate the work in order to reduce the time cost of the mission. All this infrastructure communicates under an independent ad hoc network developed by the industrial partner Humanitas Solutions allowing deployment without the use of internet or mobile network. To lead this swarm, we propose a combination of an AR HMD to assist human supervision and a touch-sensitive tablet acting as a mission planner.

\begin{figure}
	\centerline{\includegraphics[width=0.9\linewidth]{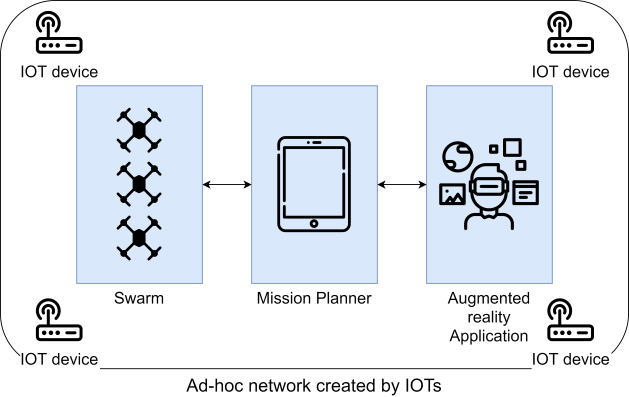}}
	\caption{Diagram representative of a use case with the system to improve situational awareness}

	\label{figure:IOTs_schema_technology}
\end{figure}

This combination of these two synchronized technologies send actions and receive information from the UAVs. On the one hand, the tablet allows the UAV to act as a mission planner from a screen, i.e. to plan mapping and building analysis missions, but also to send the objectives of these missions to the UAV, which shares the tasks independently. The mission planner also obtains information from the UAVs to inform the user of the status of each robot. On the other hand, the headset allows the user to view the data by hologram in the real environment in three dimensions. The user is able to see the information in an innovative way and positions it in the real world in order to facilitate and accelerate the consumption of information. The proposed technology consists of an Augmented Reality Mission Planner. A representative schema can be found in the figure~\ref{figure:IOTs_schema_technology}. The objective of our proposed technology is to reduce humans' confusion and to maximize user situational awareness.

\begin{figure}
	\centerline{\includegraphics[width=0.9\linewidth]{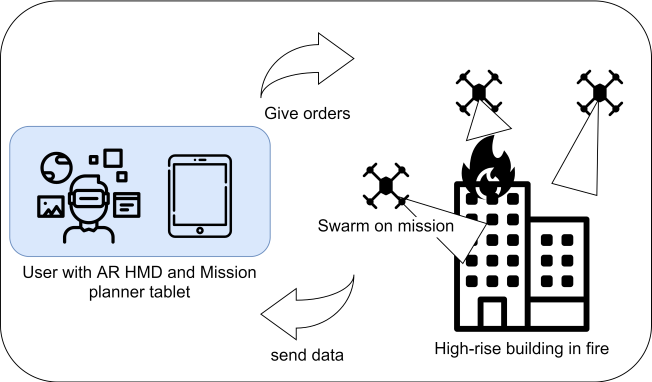}}
	\caption{Representation of the sending and receiving of information in augmented reality to the autonomous system in action}
	\label{figure:schema_resume_technology}
\end{figure}

We devised an experiment to put our technology to the test. The possible actions were divided into interactions from the headset or tablet (hands-based mainly). These interactions were necessary to consume information from the machines and to generate guidelines for the swarm. For this purpose, static, dynamic and situational tools have been developed to improve the users' situational awareness in an emergency operation. These tools responded to information needs during an emergency mission. Each tool was represented in two or three dimensions (2D/3D) depending on the type. A schematic example of a use case can be found in the figure~\ref{figure:schema_resume_technology}, the standalone system mapped a designated area to retrieve information. The person could use various actions to interact in a 3D AR universe to process information or send commands to the system. The user could interact around him with an interface, but could also visualize information in AR at the building area. He also used the mission planner on a 2D surface for more efficient actions.

\begin{table}
\begin{center}
\caption{List and function of the application widgets}
\begin{footnotesize}
\begin{tabular}{|l|p{0.6\linewidth}|}
  \hline
  \footnotesize\textbf{Name} & \footnotesize\textbf{Details}\\
  \hline
  Main menu & Main menu of the application including widgets activation and settings control to move widgets\\
  \hline
  3D Model & Display 3D models dynamically for helping user to observe the situation\\
  \hline
  Mission creation & Adding a series of waypoints to create a path for the swarm to follow\\
  \hline
  Notification & Different kinds of notification to alert the user of some events\\
  \hline
  UAV list & Allow drone selection by looking directly at or by selecting it from the list containing all drones in the swarm\\
  \hline
  UAV data & Panel containing all information about the drones, like GPS coordinate, battery level, speed, and pictures taken from the camera\\
  \hline
  Situational compass & Situational widgets helping user to locate himself in the environment containing distance and location of all drones and mission\\
  \hline
  Improved visibility & Extra visibility for swarm by colour highlighter and missions to improve tracking in the real world \\
  \hline
  Firefighters SIM form & Specific form adapted for SIM operation with fire building case\\
  \hline
\end{tabular}
\end{footnotesize}
\label{table:widgets_list_application}
\end{center}
\end{table}

\begin{figure}
	\centering \includegraphics[width=.97\linewidth]{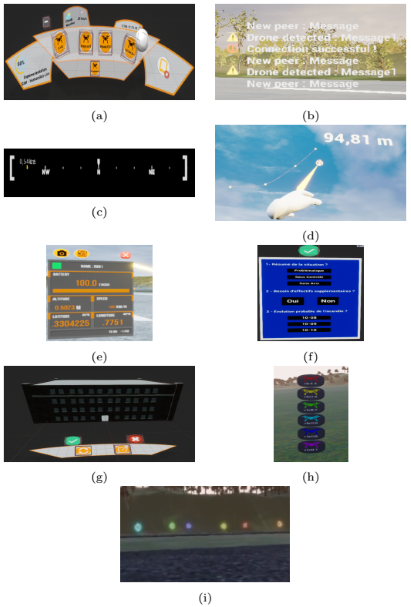}
	\caption{Image of different widgets from the application (a) Main menu (b) Notification (c) Situational compass (d) Mission creation (e) UAV data (f) SIM form (g) 3D model (h) UAV List (i) Improved Visibility}
	\label{figure:widgets_list_application}	
\end{figure}

Our application regrouped different kinds of widgets. We proposed tools listed on the table~\ref{table:widgets_list_application} as proper awareness improvement of the user. Widgets were also conceived for receiving data from the swarm and controlling it. Widgets could be placed all around the user (180 degrees in front of him). They could be easily dragged and dropped around the user using the participant's hands. We tried to imitate some mechanics from common computer actions (reduce windows, drag windows) to allow some freedom to the user in the interface and also to remind him moves that he already knew from using actual technology. The figure~\ref{figure:widgets_list_application} shows the representation of widgets. The application is directly synchronizing with the mission planner (tablet) and communicates through the ad hoc network of industrial partners allowing the system to work in every situation without 4G/5G mobile or internet. The mission planner acts as a main receiver which transmits information to the HMD. If necessary, the HMD sends command to the mission planner which transmits to the swarm.

\begin{figure}
	\centerline{\includegraphics[width=0.8\linewidth]{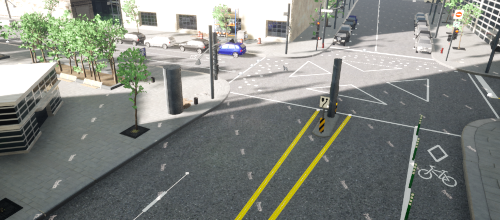}}
	\caption{The Square Victoria Square built by the Industrial Partner Humanitas Solutions}
	\label{figure:square_victoria_unreal}
 \end{figure}

\begin{figure}
 \centering \includegraphics[width=0.95\linewidth]{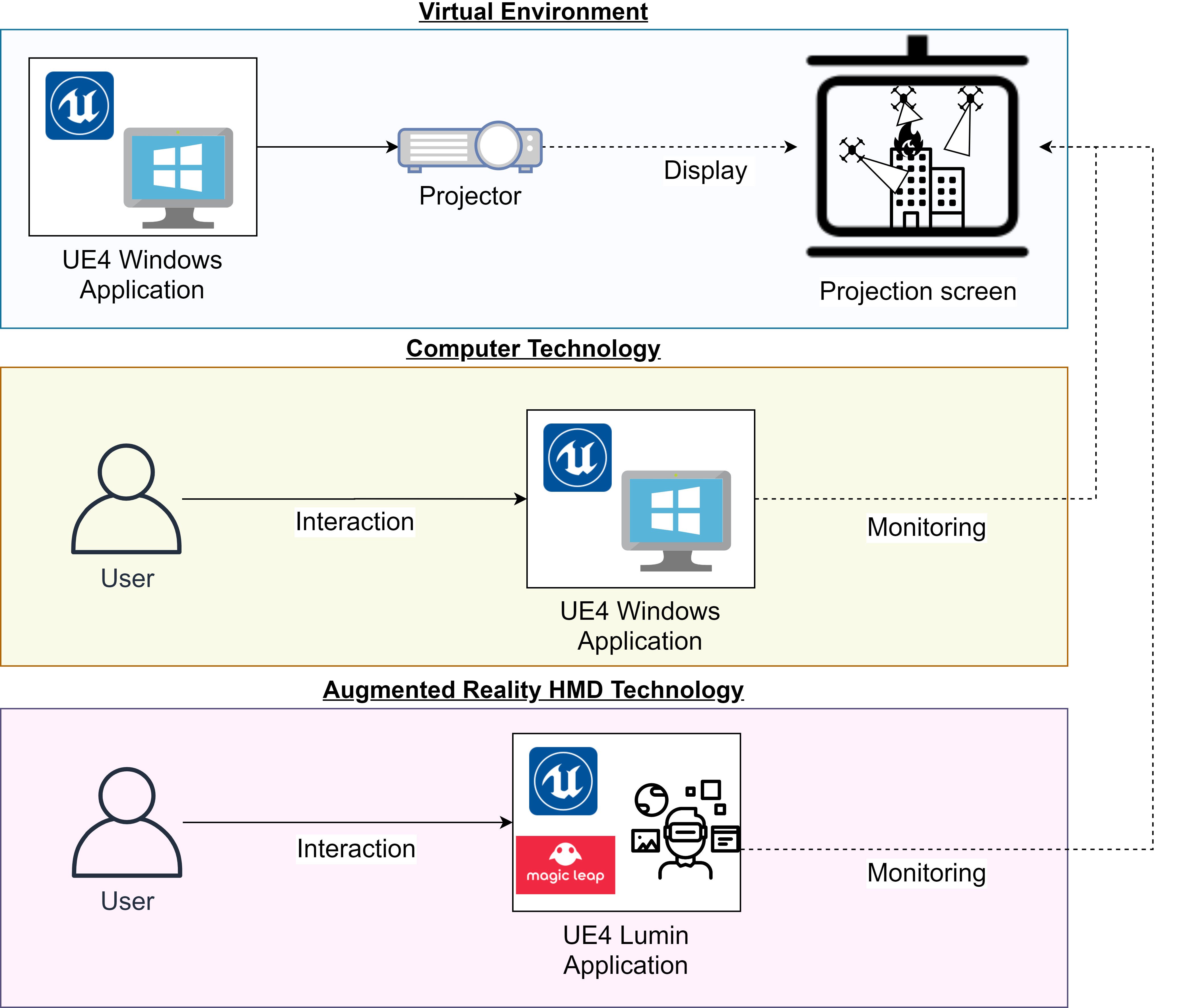}
 \caption{Complete Infrastructure of the Simulation}
 \label{figure:logistique_final_simulation}
\end{figure}

To simulate a real environment and test our work, Square Victoria in Montréal has been reproduced in a virtual 3D environment and allowed us to experiment our emergency scenario in an environment close to reality. Virtual drones were inside the virtual environment using their real GPS coordinates. A four-story high-rise building (HRB) was created to set up a fire simulation. A virtual fire started in the building, and a swarm of UAVs was available to follow the mission planner’s instructions to scour the building. The figure~\ref{figure:square_victoria_unreal} is a representation of the virtual space of the place. The figure~\ref{figure:logistique_final_simulation} contains the representation of the complete infrastructure of our experience.

Our technology is an application developed with Unreal Engine 4.24. We used the MagicLeap 1 from MagicLeap as AR HMD~\cite{magic_leap_magic_2019}. Our contribution regroups the augmented reality app and a network communication plug-in designed for the ad hoc system. The industrial partner Humanitas Solutions created the mission planner application using Unity 5. An Amazon fire tablet was used for the experience. Drones were virtually deployed as Docker-based simulation agents through Humanitas Solutions' Hyper-X-Space (HxS) rapid development platform\footnote{www.hxs.ai}. Each simulated drone is composed of i) a software-in-the-loop computing element running the embedded swarming intelligence (related to the work of Karydes and Saussié~\cite{karydes_distributed_2020} and Costa et al.~\cite{costa_covering-assignment_nodate}) and ii) a software-in-the-loop PX4 flight controller, which is defined as a peripheral element. The AR headset and the mobile mission planner could communicate with the simulated drones through a bridge powered by Humanitas Solutions' ad hoc networking technology. Every 3D virtual world was also designed by them.

\subsection{Context of the Simulation}\label{simulation_context}

\begin{figure}
	\centering \includegraphics[width=.98\linewidth]{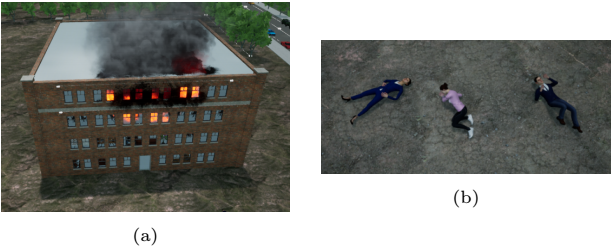}
	\caption{Pictures of virtual environment (a) HRB in fire (b) Sample of virtual people in help}
	\label{figure:virtual_building_people}
\end{figure}

To experiment our technology and to validate our assumptions required elements that may be hazardous to human health. It wasn’t possible to test our technology in real life situations during the COVID-19 pandemic. Initially, the objective of this experiment was to test the effect of AR by using a headset to direct a swarm of autonomous UAVs using a mission planner (tablet) to inspect the surroundings of a burning building to retrieve important information. Instead we used a video projector on a giant screen to display the virtual environment in figure~\ref{figure:virtual_building_people} containing critical elements of the emergency scenario involving a HRB caught by flames. The user was positioned in front of this immersive screen. Equipped with the headset and mission planner, the AR was calibrated to operate over the virtual display. For the experience, the user didn’t need to make any movements and had an overview of the situation.

Considering the COVID-19 pandemic, we adapted our experiment for testing our technology. We experimented with participants of Polytechnique Montréal University without knowledge of emergency situations. We reviewed our tests to set up a stressful situation with our technology. We virtualized a building on fire and used the swarm of UAVs to patrol around it. The participant had to manage and understand the feedback of the real-time swarm with a technology he was not familiar with over a short period of time. We removed the mission planner tablet tool and the swarm was completely automatic and did not receive any orders from the user. We mainly focus on understanding the situation using information consumption tools. To evaluate the AR performance, we also developed a similar application for 2D computer to compare the situation using the two technologies in our emergency experience.

We virtualized a HRB caught in a fire containing virtual injured people. The participant was accompanied by an autonomous system composed of 4 drones equipped with cameras patrolling the building from the outside. The challenge for the participant was to retrieve multiple information linking knowledge and location in a limited time in order to simulate a complex and stressful situation. The participant needed first to indicate the location of the source of the fire as accurately as possible. Then he or she reports the number of adults and children people seen in the building. And finally, for each kind of person, the participant needed to specify the most precise location as possible (Example: ``I saw a child on the 3rd floor of the building, he was on the northeast side of the building very close to a window'').

\begin{figure}
	\centering \includegraphics[width=.98\linewidth]{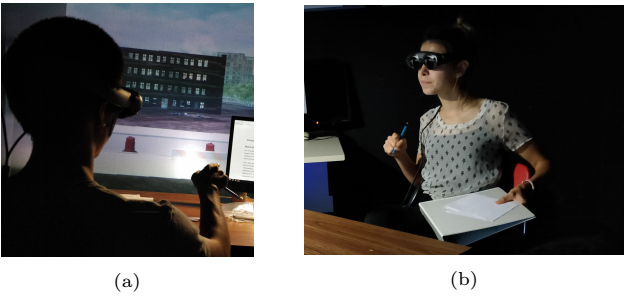}
	\caption{Experimentation with Magic Leap 1 (a) User interaction during the mission (b) Side view of the user during experience}
	\label{figure:Monitoring_software_representation}
\end{figure}

A mission had a 6-minute time limit. The participant needed to report everything he or she observed on a report to assist in a firefighter intervention. The movement of the drones from the swarm was totally autonomous in order to patrol simultaneously around the building providing the participant with several ``real time'' video feedback. In order to compare the AR performances, the user experienced two different missions and for each situation he used a different monitoring system.
From the figure \ref{figure:logistique_final_simulation}, the first one was a workstation or personal computer (PC) with a ``traditional'' 2D 24-inch monitor screen from computer receiving data in real time including videos of the drones in order to perceive the location of the drone's position. The participant used the 2D monitor and the view of the simulation environment from the video projector to improve the participant's situational awareness and report what he understood. The second was based on work with an AR headset in order to develop our studies and establish an efficiency comparison with the first system. The participant was equipped with an AR headset. As in the first case, the drones sent information and the user used the tools at his or her disposal with 3D hologram’s visual environment to understand the situation and report as much as he or she learned. For this experiment, we had two different configurations for the exercise (number and location of people, fire location) to avoid any form of repetitiveness. Half of the participants started with the AR while others start with the PC to avoid any significant order bias in the tests. The same applies to the order of the two fire configurations used. Representations of our experience are shown in the figure~\ref{figure:Monitoring_software_representation}.

Our experience was composed of three software applications developed with Unreal Engine 4.24. We used our proposed technology to create this exercise. First we made a Windows application containing the building on fire, virtual drones and virtual people according to the ``real field'' emergency environment. This was projected onto a projection screen (image (a) of figure~\ref{figure:Monitoring_software_representation}). The second app was the AR Lumin application for monitoring the swarm and contained widgets of our work. Finally, the last software application was an arrangement of the widgets we had but adapted to a windows software for PC using the mouse and keyboard. These applications communicate through a single private network.

Before each mission, a short training session was conducted beforehand to initiate the user with the software. He or she started the exercise by facing the projection screen displaying a HRB composed of 4 floors and 4 drones.

The participant didn't need to move around during the exercise. Only the virtual drones moved. As the mission started, the fire appeared and the drones started their turn. The monitoring technology received video feedback and location from each drone throughout the exercise. Each UAV circled the building from outside on a specific floor, meaning that drones were on the same position but at different altitudes. Video feedback was unique from each drone. The swarm patrolled twice around the building for 4:30 minutes and then the mission stopped. The limit duration of a mission was 6 minutes but participants were allowed to stop earlier. Then they needed to complete a mission report for each mission and at the end fill a questionnaire to get subjective opinion on the course of operations and technologies.

Both technologies had the following capabilities. Applications contained video feedback information on the status of the 4 UAVs simultaneously. A mini-map with an upside view was displayed containing positions of UAVs all around the building with cardinal points. Target buttons were available for each drone to save drone location and mark its position on a mini-map. It was possible to focus on a particular UAV by selecting it. The particularity on the PC was that all those tools were on one screen whereas widgets were displayed in front of the participant in AR. Selection was with the mouse on the PC and the index finger touch in AR. The augmented reality added a main menu and possibilities to improve visibility of UAVs and also their trajectories around the building over the projector rendering. Virtual environment and AR applications are shown on figure~\ref{figure:Monitoring_software_representation}.

\subsection{Evaluation}

The scenario in section~\ref{simulation_context} has been designed to obtain a response to the assumptions made in the section~\ref{Objectives}. First we evaluated a score depending on the results of all participants from each mission. Raw results were processed to measure the efficiency of the operations according to the technologies used and to obtain a percentage success score. Each question was evaluated according to a score indicated below. This score resulted from the following correction in the table~\ref{table:evaluation_tab}. From this table, the final score or each participant was a sum of every section for a total of 17 points converted in a percentage.

\begin{table}
\begin{center}
\caption{Evaluation of participants' feedback results during experience}
\begin{footnotesize}
\begin{tabular}{|c|p{0.8\linewidth}|}
  \hline
  \footnotesize\textbf{Score} & \footnotesize\textbf{Location of the fire outbreak}\\
  4 & Accurately indicated location (correct floor and approximate floor location)\\
  3 & Location indicated without precision (good floor only) \\
  2 & Location indicated with a precision error (good floor and bad location)\\
  1 & No indication \\
  0 & Wrong location (wrong floor) \\
  \hline
   \footnotesize\textbf{} & \footnotesize\textbf{Number of adults present in the building}\\
  4 & Exact number\\
  3 & Exact number around plus or minus 1 person \\
  2 & Incorrect number less than at least two persons\\
  1 & No indication\\
  0 & Incorrect number greater than at least two persons \\
  \hline
  \footnotesize\textbf{} & \footnotesize\textbf{Number of children present in the building}\\
  4 & Exact number\\
  3 & Exact number around plus or minus 1 person \\
  2 & Incorrect number less than at least two persons\\
  1 & No indication\\
  0 & Incorrect number greater than at least two persons \\
  \hline
  \footnotesize\textbf{} & \footnotesize\textbf{Average of each response on each person's location}\\
  4 & Precisely indicated location (correct floor and approximate location on the floor)
\\
  3 & Location indicated without precision (good floor only) \\
  2 & Location wrongly indicated but with the good floor\\
  1 & No indication\\
  0 & Wrong Location with the wrong floor \\
  +0.5 & Bonus: People correctly identified (adult or child) \\
  \hline
  \footnotesize\textbf{} & \footnotesize\textbf{Operation time }\\
  1 & Participant finished the exercise before the 6-minute time limit\\
  0 & Participant completed the exercise at the end of the 6-minute time limit.\\
  \hline
\end{tabular}
\end{footnotesize}
\label{table:evaluation_tab}
\end{center}
\end{table}

We also analyzed the subjective opinions of participants regarding their feelings on the use of the different technologies. The questionnaire in the table~\ref{table:personnalized_form} contained questions directly related to our assumptions to understand the experience of participants. Each question in the table~\ref{table:personnalized_form} was answered with values between 0 and 5, with the following meanings:

\begin{footnotesize}
  \begin{tabular}{ll}
 0 - No review        &  3 - Neutral               \\
 1 - Totally disagree &  4 - Pretty much agree.    \\
 2 - Rather disagree  &  5 - I couldn’t agree more.\\
  \end{tabular}
\end{footnotesize}

It was useful to measure the user's cognitive effort in order to understand the complexity of the tasks and to assess whether the information was correctly processed by the proposed system so that it could be easily understood. On the one hand, it helped to understand the level of difficulty experienced by the participant in terms of the physical use of the technologies. On the other hand, it improved the management of all the information and actions to be undertaken for the mental effort. Our questionnaire contained questions to also compare the comfort of both technology and especially with the experience with augmented reality systems. For each question, an additional field was present to collect comments in order to detail the answer choices. Another field was also included to collect additional opinions and comments from the participant.

\begin{table}
\begin{center}
\caption{Personalized questionnaire for our experiment}
\begin{footnotesize}
\begin{tabular}{|c|p{0.85\linewidth}|}
  \hline
  \footnotesize\textbf{1} & \footnotesize\textbf{Technical questions on the proposed Augmented Reality technology}\\
  1.1 & Interacting with the hands on the buttons in augmented reality was simple and intuitive.\\
  1.2 & Information from the drones that make up the autonomous system was relevant.\\
  1.3 & Viewing the video from the UAV cameras in augmented reality was easy.\\
  1.4 & The mini-map tool (2D map) in augmented reality was useful.\\
  1.5 & The UAV fact sheet tool was useful.\\
  1.6 & The superimposition of AR on the simulation environment enhanced a better understanding of the operation of the UAV system and the situation.\\
  1.7 & AR had allowed me to be more confident in my decisions during the exercise.\\
  \hline
  \footnotesize\textbf{2} & \footnotesize\textbf{Issues about system difficulty}\\
  2.1 & Generally speaking, AR tested for the experiment was simple to use as a first experiment.\\
  2.2 & Overall, the 2D computer station technology tested for the experiment was simple to use.\\
  2.3 & Subjectively, what technology you enjoyed most about exercising.\\
  \hline
  \footnotesize\textbf{3} & \footnotesize\textbf{Logistics Issues}\\
  3.1 & AR had not obstructed my vision in the real world when I needed it.\\
  3.2 & Wearing the headset on my head had not obstructed my movements in the real world.\\
  3.3 & Wearing the helmet on my head had not bothered or hurt me.\\
  \hline
  \footnotesize\textbf{4} & \footnotesize\textbf{Questions on Research Issues}\\
  4.1 & I found that there was a lot of information and the experience was difficult.\\
  4.2 & For a first use, I didn't have too much difficulty to adapt myself with the AR technology.\\
  4.3 & AR allowed me to understand as well or better the situation compared to the computer.\\
  4.4 & I was not confused during the experiment.\\
  4.5 & The simulation scenario was relevant.\\
  4.6 & I found the experiment interesting.\\
  \hline
\end{tabular}
\end{footnotesize}
\label{table:personnalized_form}
\end{center}
\end{table}

Hypothesis validation criteria were needed to validate our research questions. The table~\ref{table:validation_criteria} contains the set of validation criteria.

\begin{table}
\begin{center}
\caption{Criteria for hypothesis validation}
\begin{footnotesize}
\begin{tabular}{|l|l|}
  \hline
  Hypothesis & Validation criteria \\
  \hline
  H\textsubscript{a1}
  & Question 1.1 $\geq$ 3\\
  & Question 1.2 $\geq$ 3\\
  & (Question 1.3 + Question 1.4) / 2 $\geq$ 3\\
  & Question 2.1 $\geq$ 3\\
  \hline
  H\textsubscript{a2}
  & Question 1.5 $\geq$ 3\\
  & Question 3.1 $\geq$ 3\\
  & Question 3.2 $\geq$ 3\\
  & Question 3.3 $\geq$ 4\\
  & Question 4.3 $\geq$ 3\\
  \hline
  H\textsubscript{a3}
  & (Question 1.6 + Question 4.2) / 2 $\geq$ 3\\
  \hline
  H\textsubscript{a4}
  & Question 4.2 $\geq$ 3\\
  \hline
\end{tabular}
\end{footnotesize}
\label{table:validation_criteria}
\end{center}
\end{table}

\section{Results}

\begin{figure}
 \centering \includegraphics[width=.9\linewidth]{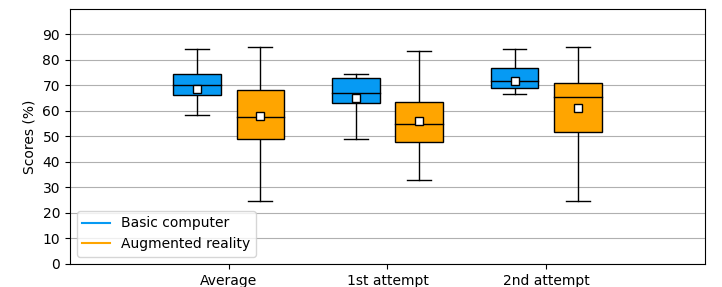}
 \caption{Box plot of missions' results by technology}
 \label{figure:results_mission_by_technology}
\end{figure}

\begin{figure}
 \centering \includegraphics[width=.9\linewidth]{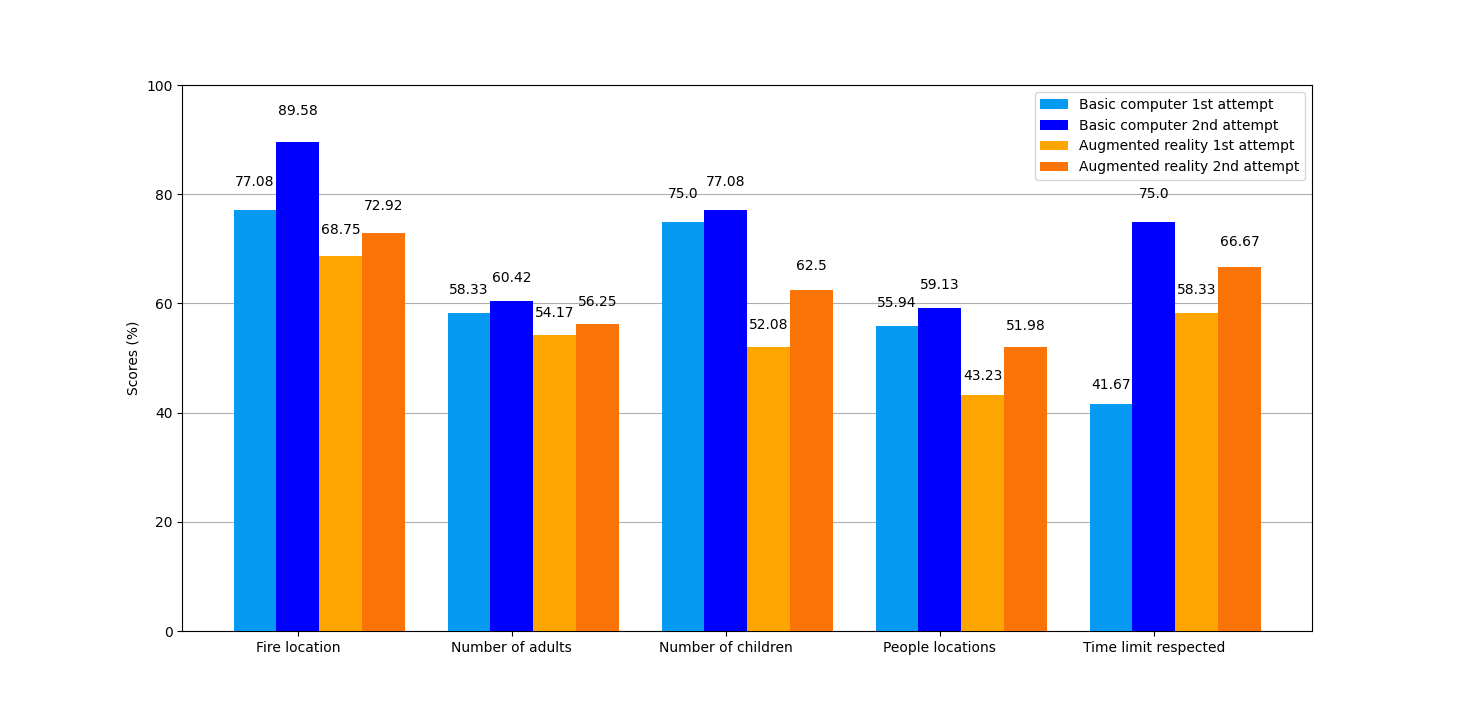}
 \caption{Details of missions report results}
 \label{figure:details_average_mission_technology}
\end{figure}

For the proposed experiment comparing PC and AR. \textit{N}~=~24 participants with no experience with AR at all and no familiarities with critical situations were asked to understand a situation where HRB took fire. The repartition of participants was 33\% of women and 67\% of men with no familiarity with AR with an HMD. They had to gather as much information as they feel possible in a short amount of time with a software application they never used before. The mission was to report multiple types of information including fire location, number of adults and children inside the building with locations for each person within a time limit of 6 minutes. We set up a stressful situation by having the participant do the exercise with unknowns, much feedback during the exercise and pressure to accomplish the mission in a short amount of time. Each participant had to complete the exercise twice. On each mission, the exercise was different (we set up two different fire scenarios). Half of the participants started with AR while the other half started with the PC to get different feedback. For each mission, they reported what they learned which results a score by the sum of each point earned depending on how right the information was (correction listed in the table~\ref{table:evaluation_tab}). The score, on a scale of 0 to 17 points is the sum of each section. We converted that into a percentage of success (0 to 100\%) for readability. The figure~\ref{figure:results_mission_by_technology} shows the average score for each technology depending on when they tried a given technology for the first time or second one. In order to measure any improvements between the use of technology depending on if it was used of first or second attempt, we computed an average for each type of information reported by participants depending on what they tried first. The figure~\ref{figure:details_average_mission_technology} represents the average score converted in percentage for each section of the mission report to analyze the improvement for both technologies and make a comparison.

\begin{figure}[h]
 \centering \includegraphics[width=.9\linewidth]{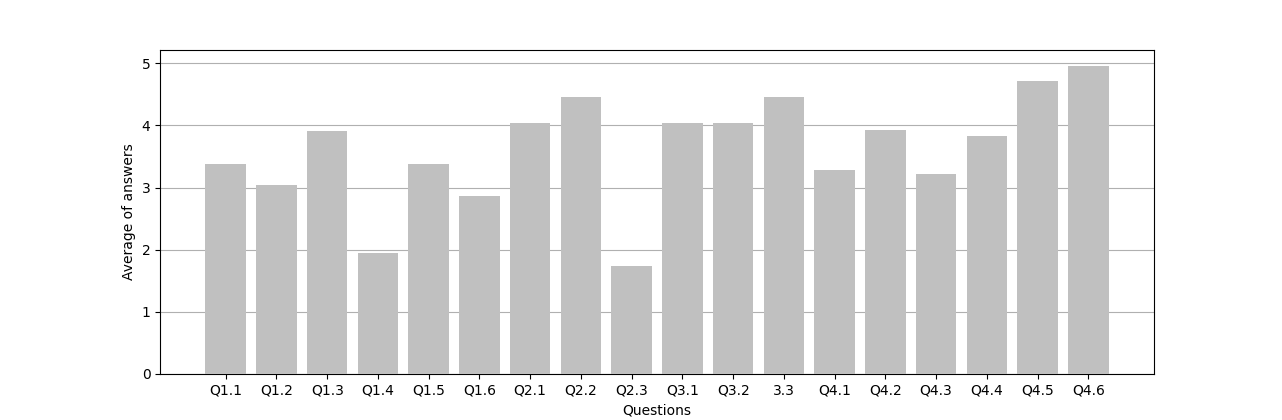}
 \caption{Responses of personalized questionnaire}
 \label{figure:personnalized_questionnaire_answers}
\end{figure}

At the end of the test, participants were asked to fill a personalized questionnaire in the table~\ref{table:personnalized_form} to have some feedback from the experience and understand their feelings. The average of all answers of these questions (between 1- strongly disagree to 5- strongly agree) is reported on the figure~\ref{figure:personnalized_questionnaire_answers}.

\section{Discussion}

We tried to measure the usefulness of AR to manage an emergency situation using a complex system. Instead of testing our work with firefighters due to COVID-19 related complications, we adapted our experience with people without experience in urgency situations.  We put the participant in a stressful situation instead of simple simulation exercise by using a short complex mission with the pressure of a strong and intense flow of information. The participant used our technology and also a PC to compare the relative performance of augmented reality. It was important to create a feeling of pressure in this situation to have a real feedback from the participant under pressure during the exercise with both technologies. It was obvious that the second attempt was to be more comfortable and less stressful than the first one, but we used that to measure the improvement of the technology after a second try and evaluated the progress between PC and AR. We also took into account the interaction method in our work in augmented reality was fully based on hand interaction without a controller and was a first-time experience for each participant.

Because each participant needed to do two exercises, two different missions were designed for the experience. Every information was different regarding the mission but the difficulty of both were equivalent. There was no important difference according to the score of each technology for each kind of mission. Both missions were used for both technologies and both attempts.

According to the boxplot figure~\ref{figure:results_mission_by_technology}, the average (white square) of the PC and AR in general is respectively $ M_{bc} = 68,44\% $  and  $ M_{ar} = 58,01\% $ showing a small difference of $ M_{bc} - M_{ar} = 10,43\% $. This figure contains the minimum ($ Q_{0}$) and maximum ($ Q_{4}$), the median($ Q_{2}$), the first and third quartile ($ Q_{1}, Q_{3} $). If we check the top 5 best scores we had among all participants, three of them were from the PC and two of them were from the AR headset with a score between $ 82\% $ and $ 89\% $. Despite the fact of a small difference, the boxplot results from PC were more stable than AR either the first or second attempt. This stability we explained by the familiarity of people with the computer. People are normally more comfortable with a technology they use every day. But for a first try with a new technology in stressful and difficult exercise, the augmented reality shows here a really good score relatively close to the efficiency of the computer. As an emergent technology, we also see that it could result in very bad score because people took too much difficulty to handle a critical situation with a tool they were not well trained for. In the same figure ~\ref{figure:results_mission_by_technology}, we notice an increase of the averages from first attempt ($ M_{bc1} = 65,12\% $ and $ M_{ar1} = 54,78\% $) to the second attempt ($ M_{bc2} = 71,75\% $ and $ M_{ar2} = 61,25\% $). Respectively, there is an increase for PC of $ M_{bc2} - M_{bc2} = 6,63\% $ and for AR of $ M_{ar2} - M_{ar1} = 6,47\% $. It makes sense by the increase of comfort generate by the experience acquired from the first attempt. We explore in detail that the increase is different regarding the technology in the next paragraph. Even if the interquartile range of PC is smaller than AR and justify the stability of the device, the minimum and maximum of boxplots are quite large for the headset resulting a good performance for some participants even in the first try. This shows the possibility that with practice and habit, the performance could be much more stable and even better than the computer.

The figure~\ref{figure:details_average_mission_technology} allows a detailed look at the improvement of each technology. These numbers are the percentage of success in each field of the mission report following the indications of the table~\ref{table:evaluation_tab}. The best improvement of the PC is shown on the fire location and time limit. The PC in blue show a good increase of $ Mfire_{bc2} - Mfire_{bc1} = 89,58 - 77,08 = 12.5\% $ compared to the augmented reality with $ Mfire_{ar2} - Mfire_{ar1} = 72,2 - 68,75 = 3,97\% $ for the fire. ``Time limit respected'' means if people finished the exercise before the limit of 6 minutes. We saw a huge improvement of $ Mtime_{bc2} - Mtime_{bc1} = 75,0 - 41,67 = 33.33\% $ compared to the headset $ Mtime_{ar2} - Mtime_{ar1} = 66,67 - 58,33 = 8,34\% $. But the average time taken by both technologies were really equal around 5:40 at the end. We noticed a similar improvement about the number of adults found ($ Madult_{bc2} - Madult_{bc1} = 2,09\% $  and $ Madult_{ar2} - Madult_{ar1} = 2,08\% $). On the last two sections (number of children and people locations), the AR showed a better improvement respectively $ Mchildren_{ar2} - Mchildren_{ar1} = 62,5 - 52,08 = 10,42\% $ and $ Mlocation_{ar2} - Mlocation_{ar1} = 51,98 - 43,23 = 8,75\% $ compared to the PC respectively $ Mchildren_{bc2} - Mchildren_{bc1} = 77,08 - 75,0 = 2,08\% $ and $ Mlocation_{bc2} - Mlocation_{bc1} = 59,13 - 55,94 = 3,19\% $. In general the PC was always better on performance for all sections here and could improve better at some point. But for a first performance on stressful and difficult missions, the augmented reality headset with new ways of interoperability demonstrated here that performance and improvement are nearly close to the computer for a very first use.

Following the missions, participants needed to answer some questions referred in the table~\ref{table:personnalized_form} to understand their feelings mainly about the usability of augmented reality. The average of answers of each question is shown in the figure~\ref{figure:personnalized_questionnaire_answers}. The first part was technical questions about AR technology. We started to ask about the interaction with the hands on the system on Q1.1, an average of $M_{PQ1.1}=3,38$ explains an optimism neutral position on that point. Some participants were really reluctant whereas some people really liked the fact to interact on buttons with their fingers which is really promising for the future. We also ask in Q1.2 how easy it was to read 2D content (a video) in hologram resulting with $M_{PQ1.2}=3,38$. Participants reported that the field of view of the headset is really small and they couldn't see well all video correctly and interact with the system at the same time whereas the computer had everything on the screen and the use of a mouse and keyboard controller are more common. The questions Q1.3, Q1.4 and Q1.5 were about widgets' utility. Q1.3 referred to the mini-map widget which was really well approved to improve situational awareness during missions with $M_{PQ1.3}=3,91$. Q1.4 wasn't considered useful at all with a poor result of $M_{PQ1.4}=1.95$. In Q1.5, overlaying information above the high-rise building and drones was a mitigated solution judged by an average of $M_{PQ1.5}=3,38$. Indeed, the most common specificity of AR is to superimpose digital information on real field. After reviewing our experiment, the participant was sitting and didn't move during the experience restricting the possibility of the functionality. However knowing that limitation proved that a result above the neutral level of 3 is not bad at all. Q1.6 ask the participant of his or her self-confidence by using AR. The answer $M_{PQ1.6}=2,86$ indicates that the question was too early after one use during the experiment and couldn't still be correctly measured at that time.
The next part is about the complexity of use about both technologies (Q2.1 referring to AR and Q2.2 to PC). Both results present a really good average with $M_{PQ2.1}=4,04$ and $M_{PQ2.2}=4.46$. The last question Q2.3 was about the most favourite technology for the experience purpose, $65,22\%$ of participants choose the PC and $34,78\%$ the AR according to $M_{PQ2.3}=1,74$. This poor result for the headset can be explained by many reasons. To accomplish a difficult mission requires a good self-confidence using a known technology. The headset was a first try to all participants and never used before which brings unknowns and uncertainties about the expected results. Participants seems to spontaneously choose the technology there were more comfortable with in order to ensure the reliability of their results for the mission. Some comments were that the augmented reality could be more favorite on the future if they have more practice with it.
The third part of this questionnaire was about the logistics of Magic Leap 1 headset. During the exercise, participants took notes while using the AR or PC. Some of them found out it was difficult to write and see with the system due to the hologram's placement in the software. The question Q3.1 approaches the vision obstruction, Q3.2 about the movement obstruction and Q3.3 on headset-related discomfort or injury from the Magic Leap 1. Every question resulted in good average: $[M_{PQ3.1},M_{PQ3.2},M_{PQ3.3}]\geq4$. It was reported that the headset wasn't well suitable for small heads and the cable could disturb a little the use of hands for interaction. But in general, the headset was really well accepted by everyone.
The last section was about the experimentation on itself. Q4.1 referred to the difficulty of missions revealing with $M_{PQ4.1}=3,29$ that participants were mitigated. It depends again on the self-confidence about the result of each participant. The next question Q4.2 asked if their situational awareness was equal or better in AR compared to the computer. $M_{PQ4.2}=3,92$ presented here an important fact that according to people, the headset was nearly equal to the computer and it's a really great proof of opportunity to demonstrate that AR could (with more use) be really accepted by people in general. Q5.3 measured the level of confusion during the test. A result of  $M_{PQ4.3}=3,22$ shows that missions were complicated but manageable. The last two questions of this form was the relevance (Q4.4) and the interest (Q4.5) of the experimentation. We conclude with $M_{PQ4.4}=4,71$ and $M_{PQ4.5}=4,96$ participants really liked the exercises and considered relevant the experimentation for the augmented reality technology in stressful conditions.

From this experience and our conditions to our assumptions, we confirm that all of our hypotheses (H\textsubscript{a1},H\textsubscript{a2},H\textsubscript{a3},H\textsubscript{a4}) are valid using our results with the exception of the condition with the mean of $M_{PQ1.3}$ and $M_{PQ1.4}$ which is just slightly below the required value. This is explained by the fact that widgets about giving GPS coordinates and information about drones weren't well relevant for the mission. We could say that AR has an opportunity to make its way into critical situations one day. We saw here that technology could provide good results in urgency operations and provide real-time information in a real situation instead of managing it all from a monitoring station. However, this technology is too emergent for the moment and needs more applications and tests to definitively prove something for the future. Our results showed that people accepted the AR technology and obtained scores really close to the PC even if PC was always better. This does not exclude the fact that with training and more comfort, the augmented reality headset could perform equally or even superior to the computer. Even in a stressful situation with an overflow of information, participants not accustomed to these conditions showed a good accommodation on this new technology with novel interactions. We also see that the improvement in AR was really close to the computer even in a first experiment which is very promising. An important complement to these tests could have been an experience with our first work or another application with people knowing the emergency field and could point out how good or bad that technology could be on a real situation and not during a simulation. Another good way for testing it would be to test this exercise on a real situation outside on a real building simulating people at risk and the fire with real drones and real video feedback. The difficulty to see holograms outside with the sunlight is still a question because holograms are made of light and become really transparent in a lighted field. But we know that technology is increasing day by day really fast and the new augmented reality headsets showed good improvements compared to the Hololens 1.0 in 2016. There is still some time until we see applications like our work used in emergency field, but we hope that this work will help to present the opportunity of augmented reality usage into critical situations.

\section{Conclusion}

In this project, we analyzed conditions of Montréal’s firefighters and reported about how new technology using augmented reality could improve humans in critical situations. After the objective’s definitions, we designed and developed an augmented reality application as an interface for an AR headset to improve situational awareness of humans. This interface allowed the user to understand information provided by a swarm of drones with 2D and 3D widgets in a virtual environment and also allowed him or her to control a swarm of drones easily to inspect external areas. To test our work, we set up a simulation with virtual environment and augmented reality in a stressful situation. Our results show that augmented reality has potential to improve the efficiency of humans and improve the success of critical missions.

\section{Acknowledgments}

This work was supported by Mitacs [grant number IT10647] and Humanitas Solutions.
It couldn’t be achieved without Computer Graphics and Virtual Reality Laboratory (LIRV) from Polytechnique Montréal. We also thank Humanitas Solutions team for all their work and support for this project. Icon made by Vitaly Gorbachev, Freepik, Nikita Golubev, Pixelmeetup from www.flaticon.com.

\begin{scriptsize}
\bibliography{references}

\end{scriptsize}

\end{document}